# Sticking behavior and transformation of tin droplets on silicon wafers and multilayer-coated mirrors

## Norbert Böwering[a]

Molecular and Surface Physics, Bielefeld University, 33615 Bielefeld, Germany, and BökoTech, Ringstr. 21, 33619 Bielefeld, Germany

## Christian Meier

Molecular and Surface Physics, Bielefeld University, 33615 Bielefeld, Germany

a) Electronic mail: boewering@physik.uni-bielefeld.de

## Abstract

Silicon wafer and multilayer-coated mirror samples were exposed to impact of drops of molten tin to examine the adhesion behavior and cleaning possibilities. The sticking of tin droplets to horizontal substrates was examined for different surface conditions in a high vacuum chamber. Silicon wafers without a coating, with thick oxide top layer, and also with differently capped Mo/Si multilayer coatings optimized for reflection at a wavelength of 13.5 nm were exposed to tin dripping. Dependent on substrate temperature and coating, adhesion as well as detachment with self-peeling and self-contraction of spreaded drops was observed. The adhesion strength of solidified tin splats decreased strongly with decreasing substrate temperature. Non-sticking surface conditions could be generated by substrate super-cooling. The morphology of non-sticking tin droplets was analyzed by profilometry. Adhering deposits were converted in-situ via induction of tin pest by infection with gray tin powder and cooling of the samples. The phase transition was recorded by photographic imaging. It caused material embrittlement and detachment after structural transformation within several hours and enabled facile removal of tin contamination without coating damage. The temperature-dependent contamination behavior of tin drops has implications for the preferred operating conditions of extreme ultraviolet light sources with collection optics exposed to tin debris.

# 1 Introduction

Liquid drops hitting, spreading and solidifying on solid surfaces give rise to a variety of phenomena that are relevant both in nature and in industry. Dependent on droplet impact conditions and substrate properties, drops can spread and freeze with or without adhesion, they can also splash or even bounce [1]. The behavior of molten metal droplets impacting and solidifying on substrates [2] is of industrial interest, for example, for thermal spray coating [3] and for the rapid printing of electrically conductive structures [4]. The process of solidification and adhesion of molten tin droplets impinging on stainless steel plates was studied in detail both experimentally and with numerical modeling for different velocities and surface conditions by Chandra and coworkers [5-8]. The sticking of tin splats is strongly influenced by heat transfer [2] and depends critically on the droplet velocity and temperature as well as the thermal properties, composition and surface roughness of the substrate. Thin surface oxide layers can also influence the adhesion behavior of metal droplets considerably [3]. During freezing after impact release stresses and effects of surface tension may built up, and flattened splats may detach from the substrate [2]. Very recently, for liquid tin droplets impinging at fairly low velocity, de Ruiter et al. reported so-called self-peeling when examining smooth, flat, horizontally oriented surfaces of various materials with different thermal conductivity and effusivity [9]. We have also examined such conditions where tin drops do not stick to the substrate due to self-peeling, and observed even a further step in some cases, namely receding and self-contraction of already spread-out pancake-like tin splats. The referenced previous studies of tin dripping were carried out in a purging gas atmosphere and at substrate temperatures of room temperature or higher. We report here an experimental analysis of tin dripping in a vacuum environment and at substrate temperatures ranging from -100 °C to 95 °C.

Our studies are motivated by an emerging new application where tin microdroplets are used as target material for laser-initiated plasma sources producing radiation at extreme ultraviolet (EUV) wavelengths of around 13.5 nm [10]. Such light sources are utilized in high-performance scanners developed for the lithographic manufacturing of next-generation semiconductor chips with integrated circuits [11]. The produced EUV light is collected and shaped by mirrors with multilayer (ML) coatings optimized for Bragg reflection at near normal incidence [12]. Inherently, these light sources produce also tin debris which may accumulate and lead to severe contamination of collection and illumination optics causing significant reflectance losses on mirror surfaces. Deposition may occur by incident tin vapor or by macroscopic droplets and even millimeter-size drops of molten tin falling down from chamber walls and adhering to the surface of optical elements after impact [10]. Optics deterioration is reduced by various debris mitigation schemes employed inside of the source modules, mainly by means of flowing hydrogen gas and by internal protection hardware [10, 13], by magnetic deflection of tin ions [14] and by protective ML coatings with cap layers [10, 15]. Nevertheless, contamination by accumulated tin deposits can built up over time, in particular on the ML-coated collector mirror that is located near the bottom of the source chamber in very close vicinity to the hot tin plasma. Maintaining high levels of reflectivity of the EUV optics requires removal of accumulated Sn contamination in order to meet the productivity targets of the scanner. This task remains challenging during continuous operation of the laser plasma at repetition frequencies of ~50 kHz or higher for chip production, especially, when massive tin splats are deposited. The adhesion strength of tin deposits is particularly important for cleaning processes, since removal of stongly adhering contamination must not result in any coating damage caused by concurrent full or partial coating layer tear-off. In-situ tin cleaning techniques that can be performed inside of the source chamber based on slow etching of adsorbed tin by hydrogen radicals were recently developed and improved further [13, 16-18]. Steady etching can reduce the thickness of thin tin deposits on EUV optics and is of interest in order to avoid lengthy collector module swaps and related pump down times.

Previously, an embrittlement process for solidified tin drops resulting from induction of a phase transformation was described by one of us [19]. Based on this scheme, we have recently reported first tests of an alternative in-situ optics cleaning scheme carried out in a vacuum chamber where massive tin drop contamination was examined on mirror samples coated with uncapped Mo/Si multilayers [20]. Complete disintegration of tin drops could be initiated by means of induction of tin pest when cooling the substrates to negative Celsius temperatures [21, 22]. In this process, tin in its normal metallic phase, β-Sn, undergoes an allotropic phase change to a lower-density semiconducting phase, α-Sn, which is stable at temperatures below 13 °C [21]. The transformation occurs by a process of nucleation and growth that is induced by bringing seed crystals of α-Sn in contact with metallic tin [19-23]. Tin deposits on ML optics crack and detach during the transition, since the new phase takes up considerably more volume. Subsequently, the loose pieces of transformed tin can easily be removed from the surface. We have already verified that multilayer Mo/Si-coated mirror samples do not undergo any substantial degradation in EUV reflectance at 13.5 nm by this process [20]. Using a refined transformation procedure, we now investigate samples with capped EUV ML-coating layers directly relevant for collector mirrors in plasma light sources and also examine if there is any influence of the substrate coating composition.

During our previous studies we have made the observation that the tin droplets often did not stick to the ML-coated sample surfaces which were held at room temperature. Sticking of impinging droplets was mainly observed when the splat also contacted the holder or the edge of the sample, but usually not or at least not strongly, when the drop was only touching the very smooth surface of the mirror. For cleaning purposes it is of high importance to fully understand the conditions for adhesion or delamination of tin splats on optical surfaces relevant for reflection of EUV light. Therefore, we investigate and compare here in detail the sticking and transformation of thick drops of molten tin after incidence in free fall on various flat samples, namely on silicon wafers and mirrors coated with differently capped multilayers. Particular emphasis is put on the analysis of the influence of the surface temperature of the substrates with respect to droplet adhesion. To our previously described apparatus [20] we have added the option to heat the sample holder during tin dripping in order to mimic the heating of the collector surface by the hot plasma in a commercial EUV source. On the other hand, by cooling the substrates to negative Celsius temperatures we now explore a parameter range that has not been investigated before in the context of tin dripping. In addition, we have provided for a possibility to verify the sticking or non-sticking of tin drops to the substrates inside of the vacuum chamber. For comparison of the sticking behavior we also carried out some corresponding tests for tin dripping in air on the same samples.

Self-peeling of droplets is expected to be less pronounced on oxidic materials [9]. To analyze the dependence of the adhesion of tin drops on the layer material, we study bare and strongly oxidized silicon wafers as well as Mo/Si coated mirrors without cap layer and with protective zirconium nitride (ZrN) or zirconium oxide ($ZrO_2$) cap layers. ZrN and $ZrO_2$ are prospective protective cap layers for the coating of EUV collector mirrors [24] that are exposed in a harsh plasma environment to hydrogen radicals and tin tetra-hydride ($SnH_4$) molecules formed after dissociation of the hydrogen buffer gas that is present in the source chamber. Decomposition of $SnH_4$ molecules is inhibited on $ZrO_2$ layers [25], in contrast to metallic surfaces. The deposition of tin on these oxide cap layers is therefore reduced. ZrN cap layers are similarly stable in the environment of hydrogen radicals and very resistant to hydrogen-induced coating blistering [26] since they also form effective barriers for hydrogen diffusion [16]. This motivated our interest to compare both the sticking behavior of thick tin drops and the subsequent transformation of adhering deposits to α-Sn on bare as well as on ML-coated silicon wafers, without and with these particular cap layers, and at specific surface temperature conditions.

# 2 Experimental

## 2.1 Materials

Tin from a rod of very high purity (99.999%) supplied by Honeywell was used in the dripping experiments under vacuum. Analysis provided by the supplier for this type of Sn material indicated that the combined contribution of trace impurity elements Pb, Sb and Bi amounted to less than 1 ppm. For some dripping tests in air, tin granules of slightly lower purity (99.99%) with mass of ~125 mg each were used. As material for the heatible trip tray for the melting and dripping of drops we had previously used copper and aluminum. However, adhesion of the molten tin to the tray surface occurred often, and possible small admixtures of impurities from the hot tray material to the tin drops could not be excluded. Since impurities may potentially slow down the phase transformation we now switched to a slotted trip tray made of molybdenum which also has good heat conductivity. For this refractory material we observe no alloying and almost no wetting of the tray surface with tin during heating in vacuum. The tray could be heated up to ~460 °C by a resistive electrical heater; its temperature was monitored by a thermocouple wire (chromel-alumel type K). Small-size seed powder of gray tin for initiation of nucleation on contaminating deposits was prepared by temperature cycling of high-purity pieces of tin with repeated transformation between the α-Sn and β-Sn phase enabled by prolonged cooling to -25 °C and short heating to above 60 °C [19].

## 2.2 Setup and vacuum equipment

Our setup consisted of an ultrahigh vacuum (UHV) vacuum chamber with a preparation region for tin dripping in its upper section and a sample holder with specimen cooling and heating capability in its lower section. The base pressure of the unbaked vacuum chamber was $10^{-5}$ Pa. Tin dripping experiments were typically conducted at vacuum pressures of around $10^{-4}$ Pa. Before pump-down several pieces of tin (mass each between ~100 mg and ~170 mg) were placed on an insertible Mo trip tray with two beveled slots (2 mm width) for dripping. The pieces melted in vacuum during slow heating of the tray to temperatures above the melting point of Sn and generally contracted while on the tray into elongated or round balls at tray temperatures of about 240 – 250 °C before dripping through the slots at slightly higher temperatures [20]. For spheres of liquid tin this resulted in droplet diameters in the range of 3.0 – 3.6 mm. The drops hit the flat samples mounted horizontally on a copper cooling block at a distance of 41 cm below (corresponding droplet speed: 2.84 m/s). They created circular solidified tin splats with diameters typically in the range of about 8 – 11 mm. The generated tin deposits could be infected with gray tin powder by sprinkling seed particles onto them from above by emptying two small receptacles attached to an insertible manipulator rod [20]. Optical monitoring of tin contamination at regular intervals was accomplished from the top through a vacuum window using a digital camera with zoom objective (Canon EOS 350D).

Several changes and additions were made to the previously described equipment [20]. A schematic view of the modified configuration is shown in Fig. 1. For improved sample illumination light from a commercial panel consisting of light emitting diodes and front diffuser was deflected into the chamber via a semitransparent acrylic plate mounted in front of the camera lens at an angle of 45 degrees. A stiff wire attached to a wobble stick mounted in the lower section of the vacuum chamber could be inserted to reach with its tip the tin splat deposits on the samples. With it, the sticking of the tin drops could be tested by applying a small pushing force of up to about ~1 N parallel to the sample surface. Adhesion was correspondingly classified as weak or (close to) zero when the tin deposits could be moved easily on the surface in this way and as strong when they could not be moved at all.

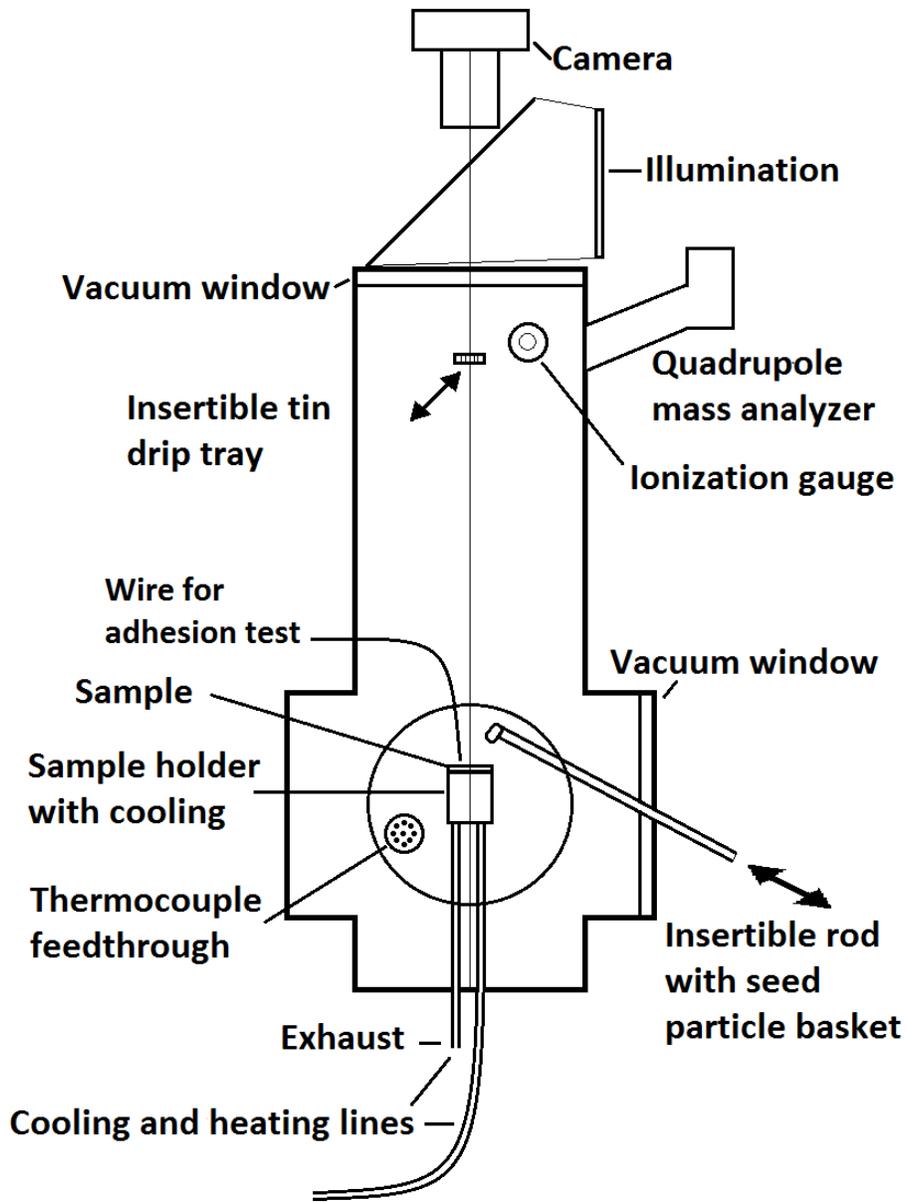

Fig. 1 Schematic view of the experimental setup. The vacuum chamber contained the equipment for tin dripping, sample infection, substrate cooling and heating, and for temperature measurements. Illumination and photographic recording was provided via vacuum windows

## 2.3 Sample holder, heating and cooling

The samples could be cooled by an insulated cooling line transporting cold nitrogen vapor evaporating from a liquid nitrogen container with pressure regulation to the sample holder made of a copper block and aluminum mounting plates. Screwed-on platelets were used to press the samples to the holder for good thermal contact. For phase transformation of tin the substrate temperature of the samples was typically regulated to the range of -30 °C to -40 °C [20], but lower temperatures could also be reached. For sticking tests at low substrate temperatures the cooling of the sample holder was usually adjusted to temperatures near -50 °C. For experiments where the sticking of tin droplets was studied during substrate heating, the sample surface temperature could be raised up to ~95 °C by connecting the line to hot water vapor generated by boiling of demineralized water in a bell jar container using a regulated heater plate. Several thermocouple-wire probes (chromel-alumel type K) were attached and used to monitor the temperatures near the sample surface, at positions at the body of the sample holder, and at the exhaust of the cooling/heating line outside of the chamber.

## 2.4 Samples

Different samples were exposed to tin dripping. Substrates were n-type Si(100) wafers with the surface polished to a roughness of <0.2 nm root-mean-square (rms). The sample thickness was 0.525 mm, 0.625 mm or 0.775 mm. The wafers were tested with bare surface (i. e., with only a native oxide layer of ~2 nm thickness), with the surface thermally oxidized (~1.1 µm thick oxide layer formed by heating in an oven under water vapor for 2.5 hours at a temperature of 1150 °C), and with Mo/Si ML coating (50 bilayers terminated with Si and native $SiO_2$ top layer, optimized for a peak reflectance $R_{max}$ above 0.65 at a wavelength close to 13.5 nm at near normal light incidence [27]). In addition, coated Si(100) samples with cap layers were used [28]: Mo/Si ML coating (50 bilayers) with 10 nm thick ZrN cap layer (peak reflectance $R_{max}$ = ~0.55 at 13.5 nm), Mo/Si ML coating (50 bilayers) with 3 nm thick $ZrO_2$ cap layer (peak reflectance $R_{max}$ = ~0.56 at 13.5 nm), and Si(100) wafer with 3 nm thick $ZrO_2$ cap layer. For comparison, tin dripping onto the rough unpolished backside of a silicon wafer (measured surface roughness: ~1 μm rms) and onto a smooth glass microscope slide (thickness: ~1 mm, material: soda-lime glass) was also examined. Typically, the samples had either square shape (24 mm or 25 mm edge length) or were larger, with the exposed surface in most cases limited to a diameter of 30 mm by the platelet of the sample holder.

## 3  Results and discussion

### 3.1 Impact of tin droplets and heat transfer to substrates

Typically, the droplet impact in vacuum proceeds in the following sequence: After release from the drip tray the molten tin droplets acquire spherical shape during their fall of 0.29 s duration to the sample. After impact, the drops spread out on the smooth substrate surface to a pancake shape within just a few milliseconds. The liquid tin starts to freeze at the outer edge of the splat where surface tension forces prevent further spreading. This happens fast, since the thermal contact resistance is low on a smooth surface and the corresponding heat transfer from the droplet to the substrate is fairly high. Several milliseconds later, the splat is frozen entirely, as can be concluded by comparison with experiments using high-speed photographic recording and modeling for similarly sized droplets incident with comparable parameters onto smooth stainless steel plates [5, 6]. In the cases examined in our studies the droplets are characterized by Weber numbers that are sufficiently low (range: 300 – 400) so that solidifying disk splat shapes dominate the process and splashing on the surface generally does not occur (except for the few cases where the edge of the sample or the sample holder is impacted). Finger patterns caused by Rayleigh-Taylor instabilities develop to some degree and become visible at the edge of the deposited splats, as was previuously discussed for droplet experiments with tin [6] and other fluids [29]. If the binding of liquid tin to the top layer of the substrate is only weak due to layer composition, high contact resistance and low wetting at low substrate temperature, tin may sometimes recede while still in the liquid state, in some cases even bounce and contract to a semi-spherical ball on the smooth surface under the force of surface tension. (Impact of tin droplets on textured Si substrates shows a somewhat similar behavior [30].) This occurs at many tens of milliseconds after impact, on similar timescales as previously observed for droplet recoil taking place at much higher surface temperatures, comparable to the melting temperature of tin [6].

As discussed in Ref. [9], the values of initial droplet temperature $T_d$, substrate temperature $T_s$ and thermal effusivity $e_s$ of a particular substrate mainly determine whether a tin drop sticks or does not stick to a homogeneous substrate. The thermal effusivity is defined as the square root of the product of the material's thermal conductivity k, density ρ, and heat capacity $c_p$. In a simplified model a limiting case can be considered, where droplet and surface (with respective thermal effusivity $e_d$ and $e_s$) are brought suddenly into contact. When treating both as semi-infinite bodies and neglecting the thermal contact resistance for heat transfer at the boundary, the interfacial contact temperature $T_c$ is given by [6, 9]:

$$T_c = T_s + (T_d - T_s) \, (1 + e_s/e_d)^{-1} \;, \qquad \text{with } e_i = (k \rho c_p)^{1/2} \tag{1}$$

Heat transfer from the hot drop to the colder substrate is also influenced by the contact resistance $R_c$ at the interfacial boundaries and, for coated samples, by the resistance to thermal heat conduction $R_t$ within the coating layers. For smooth bare silicon samples the heat is fairly quickly transported away from the droplet splat into the substrate, while for the other sample materials studied here the heat transport requires a longer time for leaving the contact region due to the lower heat conductance of the coatings. The relevant thermal properties of the examined materials are listed and compared in Table 1. With the exception of silicon, the investigated substrates and coatings have very similar effusivity values. In comparison to these, at given $T_s$ and $T_d$, the contact temperature $T_c$ is considerably lower for uncoated silicon substrates, thus leading to a higher tendency for droplet self-peeling. In contrast, for thick glass substrates droplet sticking is expected to dominate. For the case of the Mo/Si multilayer the heat transport is anisotropic; therefore, we have used here the value determined for the thermal conductivity normal to the ML plane [31, 32]. (Average density and heat capacity were estimated in this case from bulk values, including contributions of $MoSi_2$ interface layers.) For coated samples, both coating and silicon substrate with their respective effusivities contribute to the heat dissipation process.

| material | k (Wm$^{-1}$K$^{-1}$) | ρ (kg m$^{-3}$) | c$_p$ (J kg$^{-1}$K$^{-1}$) | e$_i$ (Ws$^{1/2}$m$^{-2}$K$^{-1}$) |
|---|---|---|---|---|
| liquid tin | 30 | 6990 | 209.3 | 6626 |
| solid tin | 67 | 7265 | 228 | 10535 |
| crystalline silicon | 150 | 2330 | 703 | 15675 |
| Mo/Si ML coating | 1.1 | 6480* | 465* | 1821 |
| thick SiO$_2$ coating | 1.3 | 2180 | 719 | 1427 |
| soda-lime glass | 1.0 | 2500 | 870 | 1475 |

Table1 Thermal properties of tin and of substrate materials (* indicates estimated values)

In the case of coated samples the assumption of a homogeneous semi-infinite body below the tin splat is no longer valid. Consequently, the coating layer thickness with its resistance to thermal conduction to the substrate influences the heat transport and the adhesion behavior. Therefore, it is illustrative to compare the thermal resistances for heat conduction R$_t$ in the coatings of the different samples examined in this study, leading to a ranking of their potential for droplet sticking. Table 2 lists the coating thickness l, the thermal conductivity k and the corresponding effective thermal conductance resistance per sample area A (neglecting contact resistances at the boundaries). The details of the listed coating layer combinations and thicknesses were determined from supplier information based on x-ray reflection results [27, 28]. The ML coatings of the samples have a thickness of 345 nm. They are terminated by a amorphous Si-layer that is affected by oxidation. The coatings lead to increased heat conduction resistances which are, however, of similar magnitude for the three ML-coated samples examined here, since the different cap layers are quite thin. In comparison to bare Si samples coated by only very thin oxide layers, the samples with multilayer coating and the one with thick oxide layer coating exhibit a considerably higher thermal resistance (lower thermal conductance, leading to reduced heat transfer) through the separating coating layers between tin deposits and Si-wafer substrate. The interfacial contact temperature $T_c$ directly below the tin splat is thus higher, and the tendency of the tin splat to stick and adhere to the surface is expected to be correspondingly higher for these samples.

| coating layer combination | thickness l (nm) | k (W m$^{-1}$K$^{-1}$) | $R_t$/A (m$^2$K GW$^{-1}$) |
|---|---|---|---|
| Si-SiO$_2$ | 2 | 1.3 | 1.5 |
| Si-SiO$_2$-ZrO$_2$ | 2 − 3 | 1.3 − 0.8 | 5.3 |
| Si-[Mo/Si]50-Si-SiO$_2$ | 345 − 4 − 2 | 1.1 − 1 − 1.3 | 319 |
| Si-[Mo/Si]50-Si-ZrN | 345 − 4 − 10 | 1.1 − 1 − 47 | 318 |
| Si-[Mo/Si]50-Si-SiO$_2$-ZrO$_2$ | 345 − 4 − 1.5 − 3 | 1.1 − 1 − 1.3 − 0.8 | 323 |
| Si-SiO$_2$ | ~1100 | 1.3 | 846 |

Table 2  Coating layer combination, coating layer thickness, thermal conductivity values and calculated effective thermal resistance per unit area for respective coating layers of investigated Si-wafer samples

In addition to heat conduction by the substrate, the bonding to the surface layer can influence the wetting and sticking behavior of tin drops. Compared to SiO$_2$, barrier cap layers of ZrO$_2$ are expected to show lower adhesion strength and wetting for tin. Compared to other common metals, tin has a fairly low value for the work of adhesion on $ZrO_2$ with only van der Waals forces acting, and without the formation of chemical bonds, leading to rather large contact angles and low wettability [33]. However, we did not attempt to measure the contact angles of tin drops on differently coated substrates in our experiments.

## 3.2   Tin splat morphology

Time-resolved studies of tin droplet impact using high-speed image recording in nitrogen atmosphere have already shown that droplet splats can peel off on silicon substrates after spreading [9]. Generally speaking, this occurs if the surface and interface temperatures are not very high, and low in comparison to the melting temperature of tin (231.9 °C). The self-peeling behavior is caused by thermally induced bending stress that overcomes the adhesion of the tin splat to the surface. It depends on the values of interfacial contact temperature and substrate effusivity. As a result of bending stress, at its edge the solidified droplet splat is bent away from the sample. When examining its bottom side, the splat exhibits a rich texture of grooves, corresponding to micrometer-sized annular ridges which were previously attributed to trapped air [9].

Remarkably, we observe very similar features and behavior also in our studies of tin dripping in a vacuum environment where no air can be trapped during droplet impact. Fig. 2 shows a photo of the top and two of the bottom surface (taken with different cameras and illumination) of a self-peeled tin splat deposited on a silicon wafer substrate coated with Mo/Si ML and $ZrO_2$ cap layer, super-cooled to a temperature of -50 °C. Generally, for all examined sample materials other than glass we find that the tin splats do not stick to the substrate when it is held at this low temperature. For the case of Fig. 2, a mass of 0.128 g was measured for the tin piece by weighing before the start of the drip experiment. This corresponds to a spherical diameter $D_0$ of 3.27 mm for the tin drop. Since the final splat size $D_s$ was measured to be 9.5 mm, the resulting diameter ratio, the so-called droplet spread factor [5, 6], is 2.9 in this case.

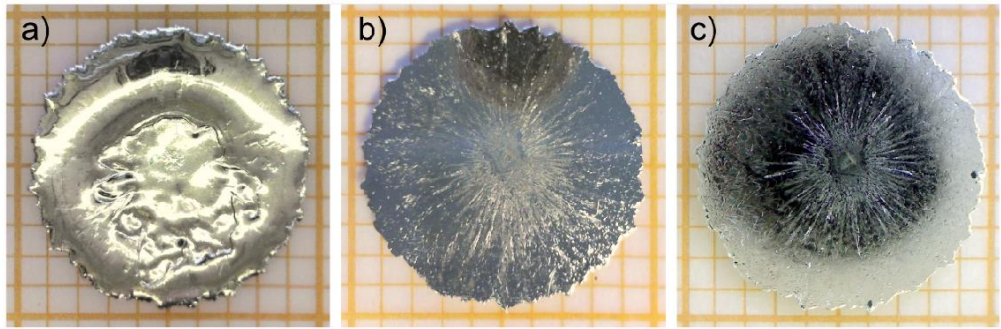

Fig. 2 Top (a) and bottom (b, c) views of a non-sticking tin splat deposited on a silicon substrate coated with Mo/Si-ZrO$_2$. A square size of 10 mm x 10 mm is indicated by the background graph paper

The droplet impact conditions are such that during arrest of spreading the solidified splat acquires a slightly higher ridge near its edge (~0.4 mm total thickness) compared to the typical thickness of the central region of ~0.2 mm. A characteristic finger pattern appears at the very edge. The bottom view shows clearly the radial pattern originating from the splat center that corresponds to these fingers. The photos of Fig. 2 are quite comparable to the splat photos of Shakeri and Chandra [34] for the case of tin drops on their smoothest steel surface. Upon careful examination a small central bump is visible in the bottom view at the very center of the drop in Fig. 2. This cavitation feature, marking the center point of impact, was observed consistently on all solidified self-peeled droplet splats obtained by us, in agreement with images of previous studies [9, 34] for tin drops and for other fluids [29, 35, 36]. Furthermore, the outer portion of the bottom side of the peeled splat exhibits an extended series of micrometer-sized annular fissures which show undulations (related in turn to the slightly higher expansion velocity in the regions where the fingers develop).

Photos of reflecting drops taken with magnification and challenging illumination do not allow to discriminate clearly between ridges, terraces and trenches on the surfaces of the splats. Using a profilometer (Veeco Dektak 3030-ST), we have therefore examined and measured the bottom side of self-peeled tin splats. Fig. 3a shows lower-resolution profile scans of entire drops, dripped on a $ZrO_2$-capped and a ZrN-capped ML sample, respectively, resulting in (slightly asymmetric) near-spherical curvatures. Curvatures in the range of 5 $m^{-1}$ to 19 $m^{-1}$ were determined for non-sticking tin splats. Fig. 3b displays a (horizontally adjusted) detail scan through a fissure region near the circumference of a droplet exhibiting 2 – 20 µm wide grooves with depths of ~300 – 1300 nm. The annular fissures likely occur as a result of bending strain in the rapidly solidifying splats that is not fully accommodated by the bulk material during freezing, resulting in crack development (for the corresponding temporal evolution, see movies of Ref. [9]). Fig. 3c shows scans through the center region of the splat photographed in Fig. 2 and for a drop dripped on a ZrN-capped ML sample. Typically, we observed depths of 3 – 5 µm and widths of about ~0.5 mm for the central cavitation. The time-resolved data of de Ruiter at el. [9] seem to indicate that this center structure appears already within the first few tens of microseconds after droplet impact. Its creation at the central stagnation point and potential causes for its generation were discussed in previous studies for different viscous fluids [29, 35, 36].

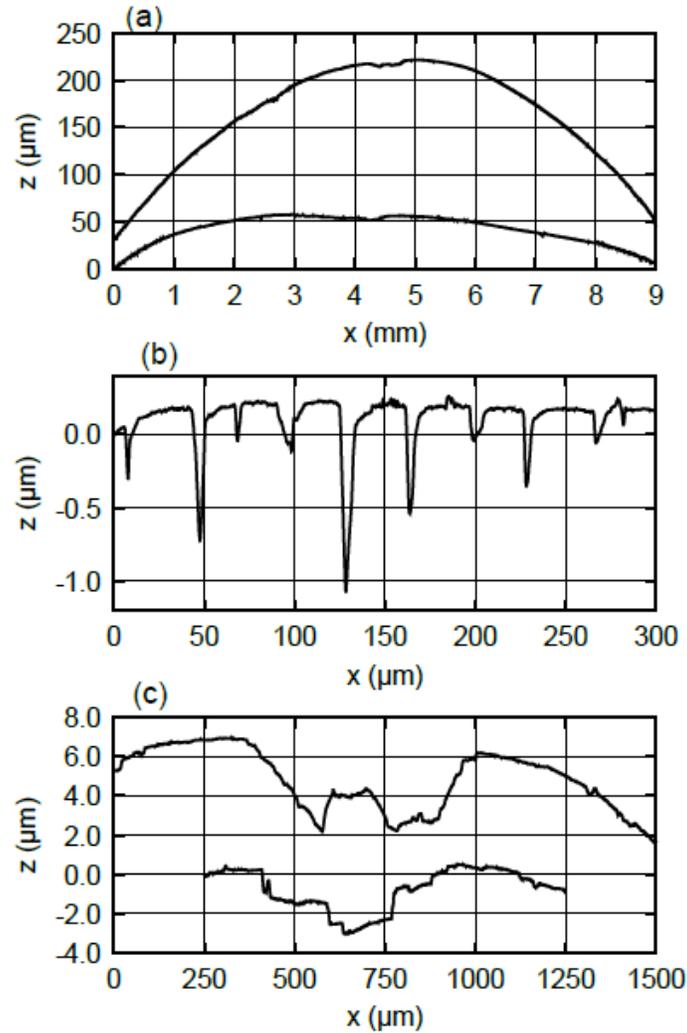

Fig. 3 (a) Profile scans of tin splat dripped on ZrO$_2$-capped and ZrN-capped substrates (upper and lower curve, respectively). (b) Scan through fissure region of tin splat. (c) Profile scans through center regions of solidified droplets (upper curve: ZrO$_2$-capped sample, lower curve: ZrN-capped sample)

## 3.2    Delamination behavior of non-sticking tin droplets

For tin dripping experiments at different substrate temperatures in vacuum we have generally investigated sample temperatures of -50 °C, 23 °C and of around ~90 °C. By use of the insertible stiff wire we determined the adhesion behavior of the splats in each case. With the exception of the rough Si wafer, the dropelts were found to stick well on all examined substrates at surface temperatures of ~90 °C. On the other hand, for surface temperatures of -50 °C, the tin splats did not stick, except for on the glass substrate, where droplet sticking was found to take place even at a temperature of -100 °C. For dripping at room temperature, sticking was observed for the Si-wafer with thick oxide layer, for the glass substrate and, weakly, for the Mo/Si-coated substrates with $ZrO_2$ and ZrN cap layers. On the other substrates (bare Si-wafer, uncapped Mo/Si ML coated wafer, Si wafer coated with thin $ZrO_2$ layer, rough Si-wafer) generally no or no strong sticking was found at surface temperatures of 23 °C. An overview of the observed sticking and peeling/contracting behavior for the examined substrates is given in Table 3.

| Sample type / $T_s$ | ~90 °C | 23 °C | -50 °C | -100 °C |
|---|---|---|---|---|
| Si-wafer | sticks | peels | peels | - |
| Si-$ZrO_2$ | sticks | peels, contracts | contracts | - |
| Si-Mo/Si | sticks | peels | contracts | - |
| Si-Mo/Si-ZrN | sticks | sticks or peels | peels | - |
| Si-Mo/Si-$ZrO_2$ | sticks | sticks | peels | peels |
| Oxidized Si-wafer | sticks | sticks | peels | - |
| Glass slide | sticks | sticks | sticks | sticks |
| Rough Si-wafer | peels | peels, contracts | contracts | - |

Table 3 Overview of sticking and detachment (peeling as well as contracting) behavior of tin drops with $T_d = \sim 250$ °C for examined samples and corresponding substrate temperatures

Although derived from idealized concepts, the limiting contact temperatures are useful reference numbers for the characterization and understanding of the droplet sticking behavior. According to Eq. (1) the contact temperatures $T_c$ for a droplet temperature of $T_d$ = 250 °C and a substrate temperature of $T_s$ = 90 °C are 221 °C for the case of glass and 138 °C for a silicon substrate. Tin drops were observed to stick in both cases. For $T_s$ = 23 °C, however, the corresponding values of $T_c$ are calculated as 209 °C for glass and only 90 °C for silicon. Under these conditions, incident liquid tin drops showed self-peeling on Si wafers during solidification. For glass at $T_s$ = -100 °C, the contact temperature $T_c$ is still 187 °C, explaining that tin droplets were found to stick also in this case. When a drop was heated on the trip tray to a fairly high temperature of $T_d$ = ~415 °C before dripping on a silicon substrate held at room temperature, it also showed higher adhesion strength leading to sticking to the surface at $T_s$ = 23 °C. Eq. (1) then results in a contact temperature of $T_c$ = ~139 °C. A similar behavior was seen when dripping such hot tin drops on Mo/Si-coated wafers. However, since the approximation of a uniform semi-infinite solid is then not valid, the thermal resistance of the coating has to be taken into account. For differently coated substrates, the trend for sticking and detachment behavior of drops with $T_d$ = ~250 °C was found to be generally consistent with the sample ranking based on thermal resistance per unit area, as listed in Table 2. Samples with higher thermal resistance $R_t$ of the coating had a disposition to show stronger adhesion for tin splats. ML-coated samples with $ZrO_2$ cap layer and oxidized Si-wafer samples consistently exhibited sticking also at substrate temperatures of 23 °C.

In a few cases, after cooling samples from ~90 °C to temperatures slightly below 0 °C, it was noticed that the splats did not stick any more for ML-coated substrates. This can be attributed to the influence of contraction strain developing during cooling due to the difference of thermal expansion properties for solid tin and Mo/Si-coated Si wafers. (The thermal expansion coefficient for Sn is more than 8 times larger compared to the one for Si.) For a direct comparison, we have also carried out some tin dripping tests in air using very similar parameters, also with a distance of 41 cm from the tray to the samples and with surface temperatures of ~90 °C and 23 °C. During heating and melting in air the tin granule is then oxidized, causing the tin drops to exhibit a non-spherical shape when dripping and leading to the appearance of scattered patches of tin oxide skin on the deposited splats. Nevertheless, for dripping tests in air we have generally observed a similar behavior of droplet sticking and self-peeling compared to dripping tests in vacuum.

Self-peeling is accompanied by slight bending of the deposited tin splats resulting in the largest distance of the drop from the sample surface to occur at large radii. If the surface tension dominates strongly over adhesion, wetting is considerably reduced and the bending forces can even lead to a reduction of the spread size. When inspecting tin dripping in vacuum on Si-$ZrO_2$ samples and on rough Si-wafers, we have observed self-contraction of already spread tin splats at times of several tens of milliseconds after impact, partially at $T_s = 23$ °C, and, even more pronounced and sometimes fully, at $T_s = -50$ °C. Partial self-contraction was also seen for tin dripping on uncapped Mo/Si-coated samples at $T_s = -50$ °C. All contracted splats were found not to stick to the substrate. Fig. 4 shows a photo of the mounted Si-ZrO$_2$ sample with two self-contracted tin drops, taken after removal of the sample holder from the chamber. Since part of the tin crown (corresponding to the tips of the fingers at the periphery) is still visible on the surface (indicated by a blue-filled circle drawn through the tin microspheres), the maximum spread size can be estimated ($D_s = \sim 8.5$ mm). This partial self-contraction was seen at $T_s = -50$ °C for a tin drop with mass of 106 mg. Interestingly, such droplet self-contraction was not observed at -50 °C for ML-coated Si samples with Mo/Si-ZrO$_2$ layers and also not for bare Si-wafers with thin native SiO$_2$ coating layer. The likely explanation is that compared to Si-ZrO$_2$ samples, the ML-coated samples with ZrO$_2$ cap layers have a higher thermal resistance for heat conduction and the SiO$_2$ terminated Si-wafers are expected to show a higher adhesion strength.

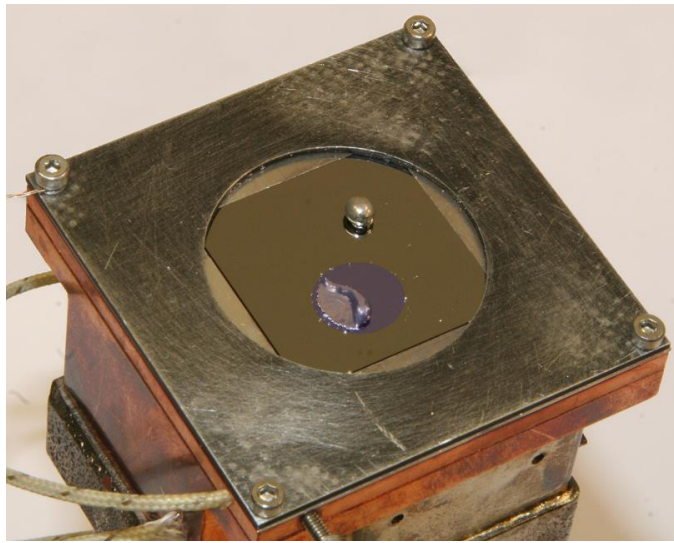

Fig. 4 Si-ZrO$_2$ sample on sample holder showing a partially self-contracted tin splat and a fully contracted drop. A superimposed blue circle serves to indicate approximately the size of the spreaded splat prior to contraction

For Si-ZrO$_2$ samples with dripping at $T_s$ = 23 °C the conditions were such that either partial self-contraction or self-peeling of tin drops was observed. However, self-peeled drops exhibited then a rather smooth top surface of very uniform thickness without any ridge at large radii, indicating that the splat stayed liquid after impact for a comparatively long time due to low adhesion. Correspondingly, by profilometry of the bottom side of such peeled drops no annular fissures were detected in the case of Si-ZrO$_2$ at $T_s$ = 23 °C. This is quite in contrast to the observed morphology of drops delaminated from the other samples. Further insight may be gained from modeling of tin fluid flow and heat transfer when the influence of the oxide layer is included. For tin dripping in air, only self-peeling and no droplet self-contraction was observed.

Peeled and contracted tin drops can easily be removed from the samples. However, for all smooth coated Si substrates held at $T_s$ = ~90 °C by heating, sticking of dripped tin drops was always found to dominate. As discussed above, the contact temperature $T_c$ is then fairly high. In comparison to Si the effusivity of reaction-bonded SiC is somewhat larger. Nevertheless, at elevated temperatures tin drops are expected to stick similarly also to smooth ML-coated SiC substrates, although such samples were not examined in our study. This corresponds to the scenario of a moderately water-cooled EUV collector mirror mounted in close vicinity to a hot laser plasma where the exposed mirror surface can easily reach average temperatures of ~60 °C or more during plasma operation at high repetition rates, even in the presence of water cooling of the substrate. Since tin drop deposits are quite adhesive at elevated temperatures and do not exhibit self-peeling, the detachment from the surface has to be induced by some other means.

## 3.3 Transformation behavior of sticking tin droplets

To achieve the goal of complete tin removal also for tin splats adhering strongly to the substrates additional measures need to be carried out. Simple cooling of the samples to slightly negative Celsius temperatures after molten tin was dripped on them at higher surface temperatures, relying on differences in thermal expansion of tin deposits and substrates, is in many cases not sufficient for achieving successful splat detachment. However, the previously explored method of induction of tin pest with subsequent in-situ phase transformation of β-Sn to α-Sn can be employed [20]. Due to the resulting large expansion of the tin deposits with a volume increase of ~25% and the concurrent material embrittlement of gray tin, the adhesion to the sample surface can be overcome by this method in a fairly short time when high-purity tin is used.

In all cases where tin splats were found to stick to the substrate at temperature $T_s$, we have carried out the transformation to α-Sn according to the following sequence: In-situ infection with gray tin powder was done under vacuum at room temperature by dropping small α-Sn seed particles onto the tin splats using the insertible manipulator rod. Next, the samples were cooled and held for several hours at temperatures in the range of -30 °C to -40 °C. Image recording of the progress of phase transformation was made by taking photos at regular intervals of 5 minutes. Cooling was stopped when the tin drops had transformed completely, as evidenced by color changes of the deposits to dark gray and by movements of loose tin pieces on the smooth substrate surfaces. These slight movements were induced by small vibrations coupled to the sample holder and originating from the rotational motion of the turbomolecular pump. Remaining gray tin pieces on the samples could easily be removed from the sample surfaces after venting. Since nucleation of gray tin at suitable sites on the surface of tin splats is a statistical process, and since the number and orientation of α-Sn seed particles varied, the time period to the onset of transformation also showed significant variations. The effective growth of gray tin areas on the deposits varied as well, since it depends on the number and location of transformation sites. This caused some temporal scatter in the observed progress of tin conversion for the various cases examined.

Fig. 5 shows a sequence of images for a tin drop dripped (at $T_s$ = 82 °C) on a Si substrate coated with Mo/Si ML and ZrN cap layer. Fig. 5a is a view of the tin splat after infection at the start of substrate cooling. A comparison with the photo of Fig. 5b, taken 78 minutes later, showed that α-Sn had started to form on the surface of the drop at several nucleation sites. Subsequently, the dark gray tin blisters grew in size during continued cooling (see Fig. 5c and Fig. 5d), leading to deformation of the droplet surface due to volume expansion. Later, cracking of α-Sn regions on the drops occured, in particular, when blisters touched each other (Fig. 5e), and the clean reflecting sample surface below became visible. The gray tin pieces also started to detach from the substrate, as evidenced by the movement of the loose α-Sn pieces on the smooth surface. Fig. 5f shows the remaining brittle pieces of gray tin after 8 hours of cooling when the transformation was complete. Correspondingly, Fig. 6 displays a similar sequence for an infected tin droplet dripped (at $T_s = 92 \,°C$) onto a Si substrate with Mo/Si ML-coating and $ZrO_2$ cap layer. In less than one hour after start of cooling (Fig. 6a) a nucleation site of α-Sn had developed (see Fig. 6b); three sites became visible after 2 hours (Fig. 6c). Subsequently, the gray tin areas grew further and large cracks developed on deformed, expanding blisters (see Fig. 6c), leading to partial disintegration of the splat (Fig. 6e) and nearly complete transformation to gray tin with detachment from the surface after about 9 hours from the start of cooling (Fig. 6f). Typical growth rates of α-Sn regions were $10 - 15 \,\mu m/min$.

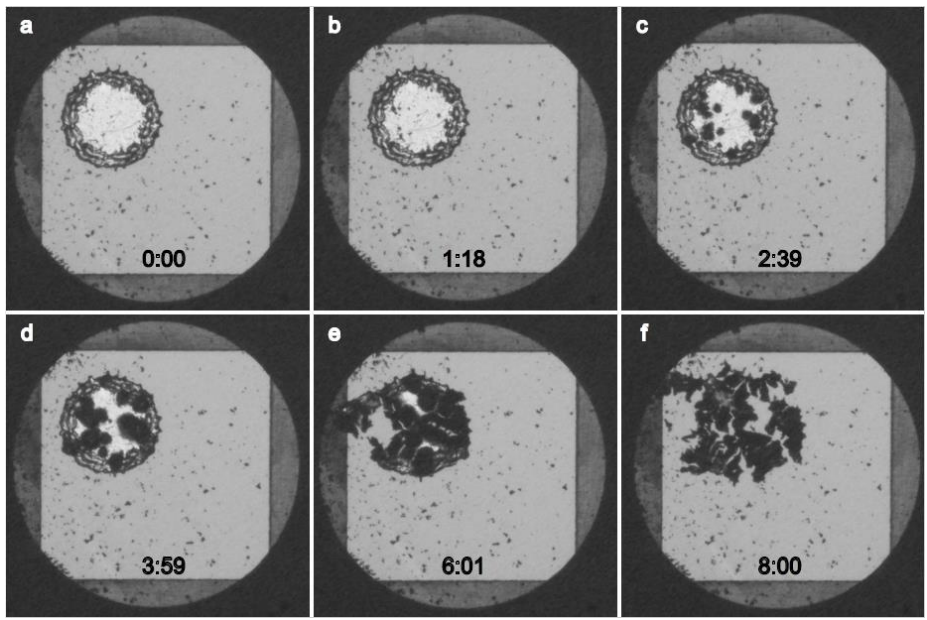

Fig. 5 Images recorded during transformation of tin drop dripped on Si-Mo/Si-ZrN sample at $T_s = 90 \,°C$. Time after start of cooling is indicated in the figure

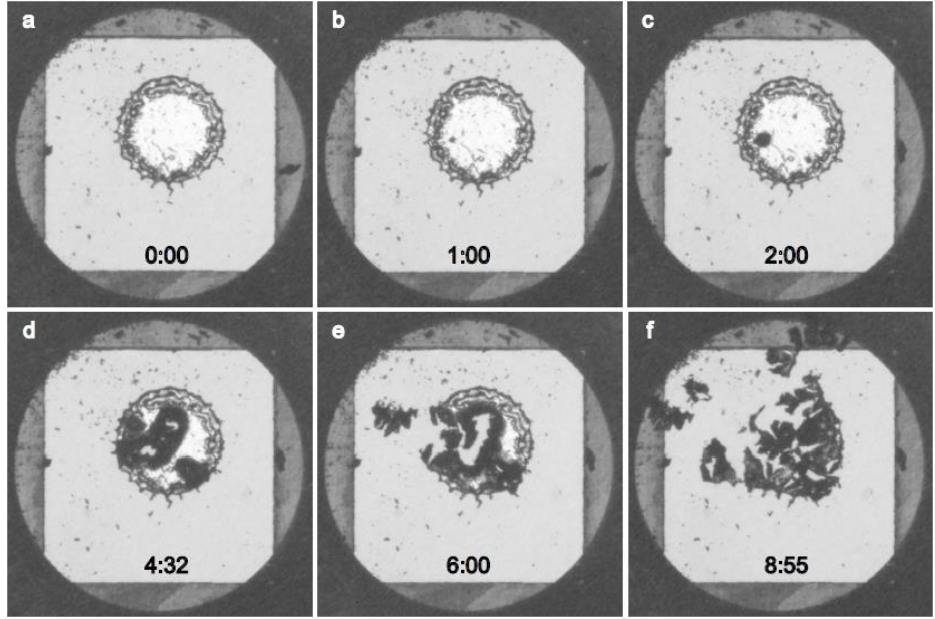

Fig. 6 Images recorded during transformation of tin drop (mass: 0.154 mg, $D_s$: ~11 mm) dripped on Si-Mo/Si-ZrO2 sample at $T_s$ = 90 °C. Time after start of cooling is indicated in the figure

The relative surface area converted to gray tin on the drops was evaluated for selected photo images of similar conversion series recorded for different samples as a function of cooling time. Fig. 7 displays these increasing converted area fractions versus time on a semi-logarithmic plot for sticking tin splats on some substrates. Tin dripping was carried out at surface temperatures $T_s$ of either 23 °C or ~90 °C. Tin dripped onto substrates with Mo/Si-ZrO$_2$ coating at $T_s$ = ~90 °C showed a typical temporal behavior with full conversion in less than 12 hours. The incubation times for nucleation of gray tin varied between extreme durations of 15 and 450 minutes. After first nucleation, the transformation to practically 100% of gray tin on white tin took in between 400 and ~1000 minutes. Full phase transition to gray tin could be achieved in less than 24 hours in all cases, and, when fast nucleation occured, often even in just a few hours. The transformation was found to be independent of substrate composition and coating, as well as history of tin deposition. It always led to a full detachment of the transformed deposits from the surface after sufficiently long duration of cooling. (In the case of the glass substrate, after in-situ transformation to ~60% coverage, the conversion was completed ex-situ to 100% in a freezer at -25 °C.) No evidence of any substrate coating damage was noticed, also after surface inspection with a microscope. The observed time variations for first-time transformed tin deposits studied here differ slightly from the behavior found in kinetic studies of tin phase transitions where multiply phase-cycled Sn material was used. There, a more regular behavior was seen, since nucleation sites were then always present and no external infection for initiation of phase transformation was required [37]. However, the data displayed in Fig. 7 exhibit the general shape that is expected for the typical time dependence of growth for such phase transformations according to the Avrami model [38].

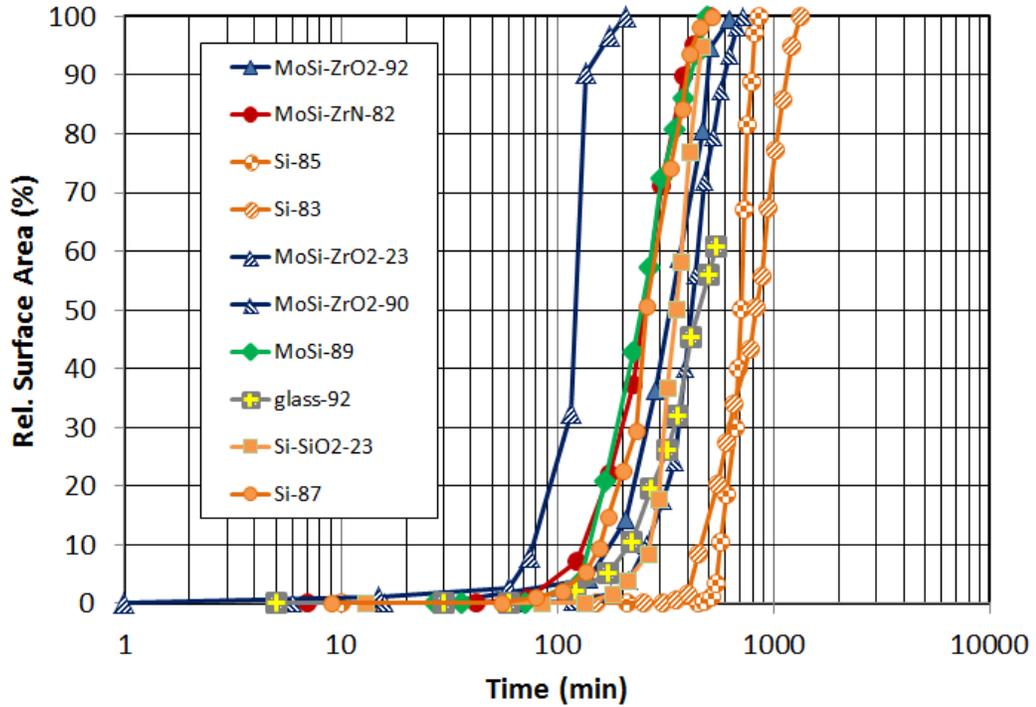

Fig. 7 Time evolution of surface area converted to α-Sn for different samples exposed to tin dripping. Brown shaded circles: Si-substrates; brown squares: Si with thick oxide coating; red circles: Si with Mo/Si-ZrN coating, blue triangles: Si with Mo/Si-ZrO$_2$ coating; green diamonds:Si with uncapped Mo/Si coating; gray squares: glass substrate. The number added to the sample type indicates the surface temperature measured during tin dripping in Celsius units

## 4. Conclusions

We have studied in detail the adhesion behavior of molten tin droplets dripped under vacuum onto smooth samples with different coatings at substrate temperatures of -50 °C, 23 °C and ~90 °C. Droplet sticking was found to be strong at surface temperatures of ~90 °C. Mainly non-sticking behavior was observed at lower temperatures on silicon substrates with different coatings. Coatings with larger thermal resistance led to an increased tendency for the tin splats to adhere to the sample surface. Droplet detachment by self-peeling or sometimes even self-contraction of deposited splats was found for non-sticking cases at room temperature and for substrates cooled to -50 °C. General considerations of heat transfer based on substrate effusivity and coating thermal resistance explain the droplet sticking and detachment behavior seen on different samples. Thicker oxide coating layers lead to a stronger tendency for droplet sticking. Profilometric measurements of the morphology of the bottom side of delaminated tin splats revealed the strong influence of surface tension causing bending and development of annular fissures in the tin deposits. Particular emphasis was put on the examination of the sticking and transformation behavior of ML-coated samples with ZrN and $ZrO_2$ cap layers which are relevant for EUV collectors. For drop deposits sticking to the various substrates, transformation to gray tin could always be induced at low temperatures after infection with $\alpha$-Sn particles, leading to embrittlement and detachment of the tin splats. In all cases tin deposits could thus successfully be removed from the sample surface. The phase transformation exhibited some variance in nucleation time and effective growth rate, but occurred similarly on all substrates. It was completed in most cases in less than 12 hours, and always within 24 hours.

Our observations have implications for the cooling conditions of collector mirrors in tin-based sources used for light generation at 13.5 nm for EUV lithography [10, 14]. In such devices the collector module is usually in close vicinity to the hot tin plasma. Consequently, during continuous source operation the mirror is exposed to a high heat load by absorption of radiation and other flux incident from the plasma. Moderate water-cooling through internal channels of collector substrates at positive Celsius temperatures likely then does not ensure that sufficiently low temperatures can be reached at the mirror surface in order to avoid the sticking of tin droplets during exposure to hot tin debris and molten tin drops. Tin deposits will therefore adhere to the mirror surface and will lead to a substantial degradation of its EUV reflectance. If substrate super-cooling with inlet temperatures below -20 °C or -30 °C would be employed instead by using, for example, a mixture of ethylene glycol and water as a cooling fluid, significantly lower contact temperatures could be reached at the surface of the collector mirrors. This would have two beneficial consequences: Sticking of tin drops (and likely also of tin micro-droplets) to the mirror surface could be substantially reduced during source operation since surface temperatures sufficiently low for the dominance of splat self-peeling conditions could then be reached and maintained. At times when the source is not operated, α-Sn microparticles could be injected in low amounts into the source chamber for infection of tin deposits. Detachment of sticking splats could then be induced by phase transformation to brittle gray tin during mirror cooling to negative Celsius temperatures. Removal of loose deposits should then be possible, greatly reducing the mirror degradation and increasing its lifetime. Both measures could lead to significant improvements of collector module up-time and overall system availability.

## Acknowledgements

This study was motivated by the ongoing industrial development of EUV light sources. We are grateful to the molecular and surface physics group at Bielefeld University for general support and for supplying Mo/Si-coated EUV mirror samples. Furthermore, we would like to thank Torsten Feigl and his team at optiX fab for generously providing several ML-coated mirror samples at our request. This research did not receive any specific grant from funding sources in the public, commercial, or not-for-profit sectors.